\documentclass[12pt]{article}

\usepackage{amsmath,amssymb}
\usepackage[english]{babel}

\newcommand{\bea}{\begin{eqnarray}}
\newcommand{\eea}{\end{eqnarray}}
\newcommand{\beano}{\begin{eqnarray*}}
\newcommand{\eeano}{\end{eqnarray*}}
\newcommand{\beq}{\begin{equation}}
\newcommand{\eeq}{\end{equation}}

\newcommand{\hs}[1]{\hspace{#1 mm}}

\newcommand{\BB}{{\mathbb B}}
\newcommand{\CC}{{\mathbb C}}

          \def\sF{\mathsf{F}}

\def\sM{\mathsf{M}}    \def\sN{\mathsf{N}}

\newcommand{\mb}[1]{\hs{4}\mbox{#1}\hs{4}}

\newcommand{\so}{\scriptscriptstyle \rm I}
\newcommand{\st}{\scriptscriptstyle \rm I\hspace{-1pt}I}

\newcommand{\la}{u}

\newcommand{\muc}{v^{\scriptscriptstyle C}}
\newcommand{\mub}{v^{\scriptscriptstyle B}}

\newcommand{\bla}{\bar u}
\newcommand{\bmu}{\bar v}
\newcommand{\blac}{\bar{u}^{\scriptscriptstyle C}}
\newcommand{\blab}{\bar{u}^{\scriptscriptstyle B}}
\newcommand{\bmuc}{\bar{v}^{\scriptscriptstyle C}}
\newcommand{\bmub}{\bar{v}^{\scriptscriptstyle B}}

\def\Izer{{\sf K}}

\def\UF{\sF}

\begin{document}
\oddsidemargin=-0.25cm.


{\large \textbf{Bethe vectors and form factors
for two-component Bose gas}}\\[1.2ex]

{ Eric Ragoucy, LAPTH (CNRS), Annecy, France}\\[1.2ex]

\textsl{Talk given at SQS'2015, JINR-BLTP, Dubna, August 3-8, 2015.}\hfill LAPTH-Conf-042/16\\

\begin{center}
\begin{minipage}{82ex}\small
This short note presents works done in collaboration with S. Pakuliak (JINR, Dubna) 
 and N. Slavnov (Steklov Math.\!\! Inst., Moscow). It is a summary of the articles 
\texttt{arXiv:1412.6037}, \texttt{arXiv:1501.07566},  \texttt{arXiv:1502.01966} and  \texttt{arXiv:1503.00546.}
Here are mentioned only references used for our calculations. A detailed list of
references can be found in our articles mentionned here.
\end{minipage}
\end{center}
%

\section{Two-component Bose gas: general context}
We consider a {one-dimensional} bose gas, with {delta interaction} and an {internal degree of freedom.}
The continuous version is described by the {Non-Linear Schr\"odinger (NLS) Hamiltonian}
$$
H_{NLS} = \int_0^L\left(\partial_x\Psi^\dagger \partial_x\Psi +c\Psi^\dagger (\Psi^\dagger \Psi) \Psi \right)\,dx,
$$
where $\Psi(x,t)=\begin{pmatrix}\psi_1(x,t)\\ \psi_2(x,t)\end{pmatrix}$ is a bosonic {two-component vector} satisfying the  canonical commutation relations
$$
[\psi_\alpha(x,t),\psi_\beta^\dagger(y,t)]=\delta_{\alpha\beta}\delta(x-y)
$$
and we assume {periodic boundary conditions} $\Psi(x+L,t)=\Psi(x,t)$ (i.e. we are on a circle).

\subsubsection*{Aims and tools}
We aim at computating the  form factors of local operators.
In particular for $x\in[0,L]$ and $j,k=1,2$  we wish to calculate
$\CC\, {\psi^\dagger_j(x)\psi_k(x)} {\BB}$,
 ${\CC} {\psi^\dagger_j(x)} {\BB}$ and ${\CC} {\psi_j(x)} {\BB}$, that we will present below.

For such a purpose, we will be in the general context of Algebraic Bethe Ansatz (ABA).
More specifically, we will use
\begin{enumerate}
\item A lattice version of the model and compute the Bethe vectors (BVs) through ABA
\item An (auxiliary) composite model to obtain a convenient expression for local operators
\item The zero mode method that helps to relate different form factors
\item The twisted transfer matrix method that allows to compute diagonal form factors
\end{enumerate}


\section{Lattice version of the Bose gas}
The lattice version  of the Bose gas was introduced  in \cite{KR1982} and its ABA developped in \cite{nikita}:
\begin{equation}\label{eq:L}
L_{{0}}(u|{n})={ \left({\begin{array}{ccc}
1+\frac{{c\Delta^2}\,\hat\sN_1({{n}})}{2-iu\Delta}
&\frac{ {c\Delta^2}\,\psi_{1}^\dagger({{n}})\psi_{2}({{n}})}{2-{iu\Delta}}
&\frac{-i\Delta\psi_{1}^\dagger({{n}}) Q({{n}})}{1-{iu\Delta}/2}
\\[1ex]
\frac{{c\Delta^2}\,\psi_{2}^\dagger({n})\psi_{1}({{n}})}{2-{iu\Delta}}
&1+\frac{{c\Delta^2}\,\hat\sN_2({{n}})}{2-{iu\Delta}}
&\frac{-i\Delta\psi_{2}^\dagger({{n}}) Q({{n}})}{1-{iu\Delta}/2}
\\[1ex]
\frac{i\Delta Q({{n}})\psi_{1}({{n}})}{1-{iu\Delta}/2}
&\frac{i\Delta Q({{n}})\psi_{2}({{n}})}{1-{iu\Delta}/2}
&\frac{2+{iu\Delta}}{2-{iu\Delta}}+
\frac{{c\Delta^2}\,\hat\rho({{n}})}{2-{iu\Delta}}
\end{array}} \right)_{{0}}}
\end{equation}
In the formula above, ${{n}}$ labels the {lattice site number} and plays the role of $x$,
while ${{0}}$ labels the {auxiliary space $V=End(\CC^3)$.}
 $\Delta$ is the lattice spacing and $u$ is the spectral parameter.
 $\Psi({{n}})=\begin{pmatrix}\psi_1({{n}})\\ \psi_2({{n}})\end{pmatrix}$
 is a discretized version of $\Psi(x,t)$ and obeys 
 $$
[\psi_{i}(n),\psi_{k}^\dagger(m)]=\frac1\Delta {\delta_{ik}\delta_{nm}}.
$$
We use the notation $\displaystyle Q({{n}})=\sqrt{c+\frac{c^2\Delta^2}4\hat\rho({{n}})}$, where
$ \hat\rho({{n}})= \hat\sN_1({{n}})+\hat\sN_2({{n}})$ is the total number of particles
and $\hat\sN_j({{n}})= \psi_{j}^\dagger({{n}})\psi_{j}({{n}})$, $j=1,2$ are the particle number operators.

The $L$-operator satisfies the {$RTT$-relation}
\beano
&&R_{12}(u,v)\,L_1(u|n)\,L_2(v|n)=L_2(v|n)\,L_1(u|n)\,R_{12}(u,v)\\
&&R_{12}(u,v)=\mathbf{I}+g(u,v)\,\mathbf{P}_{12} \mb{with} g(u,v) = \frac{-ic}{u-v}\,.
\eeano
$\mathbf{P}_{12}\in End(\CC^3)\otimes End(\CC^3)$ 
 is the permutation operator of the two auxiliary spaces 1 and 2.
$R_{12}(u,v)$ is a {GL(3)-XXX $R$-matrix} (associated to the Yangian $Y(gl_3)$).
\paragraph{The vacuum expectation values} $\psi_{j}(n)|0\rangle_n=0$, $j=1,2$, lead to
\bea\label{eq:Lvac}
\, L(u|n)\,|0\rangle_n= 
\begin{pmatrix} 
1 & 0 & \frac{-i\Delta\sqrt{c}}{1-\frac{iu\Delta}2}\, \psi_{1}^\dagger(n) \\[1.4ex]
0 & 1 & \frac{-i\Delta\sqrt{c}}{1-\frac{iu\Delta}2}\, \psi_{2}^\dagger(n) \\[1.4ex]
0& 0 & r_0(u)
\end{pmatrix} |0\rangle_n
\mb{where}
r_0(u) = \displaystyle\frac{1+\frac{iu\Delta}2}{1-\frac{iu\Delta}2}\,.
\eea


\paragraph{The monodromy matrix} reads
$
T(u)\equiv T_0(u|12\dots M)=\,L_0(u|M)\cdots L_0(u|2)\,L_0(u|1),
$
 where $M$ is the number of lattice sites. 
 Since $T(u)$ obeys the $RTT$ relation, the {transfer matrix $t(u)=tr_0T(u)$} defines an 
{integrable model} that includes  in particular a lattice version of $H_{NLS}$.

In the representation associated to
$|0\rangle=|0\rangle_1\otimes|0\rangle_2\otimes\cdots\otimes|0\rangle_M$,
  we get:
\beq
T_{jj}(u)|0\rangle=\lambda_j(u)\,|0\rangle\mb{with}
 \lambda_1(u)=\lambda_2(u)=1\mb{and} \lambda_3(u)= \big(r_0(u)\big)^M\,|0\rangle.
\eeq


\paragraph{We get back to the continuous version through the limit: $\Delta\to 0$ and $M=\frac{L}{\Delta}$.}
In the continuous limit, $\lambda_3(u)\to e^{iuL}$, which implies:
\beano
T_{ij}(u) &\to& \delta_{ij}+\frac{ic}{u}\int_0^L \psi_i^\dagger(y)\psi_j(y)\,dy+O(u^{-2}),\qquad i,j=1,2,
\\
T_{i3}(u) &\to& -\frac{\sqrt{c}}{u}\left({e^{iuL}}\psi_{i}^\dagger(L)-\psi_{i}^\dagger(0)\right)+O(u^{-2}),\qquad i=1,2,
\\
T_{3j}(u) &\to& -\frac{\sqrt{c}}{u}\left(\psi_{j}(L)-{e^{iuL}}\psi_{j}^\dagger(0)\right)+O(u^{-2}),\qquad j=1,2,
\\
 T_{33}(u) &\to& {e^{iuL}}-\frac{ic}{u}\,{e^{iuL}}\!\int_0^L
\bigl(\psi^\dagger_1(y)\psi_1(y)+\psi^\dagger_2(y)\psi_2(y)\bigr)\,dy+O(u^{-2}).
\eeano
Remark that $T(u)$ does not provide access to local operator, because $L(u)$ is not based on $R(u,v)$. Note also the
'unusual' behavior of $T(u)$ as $u\to\infty$ (w.r.t. 'usual' spin chains). Then, to compute form factors of locl operators,
we need to consider {composite models} and a {modified version of the zero mode technics.}

\section{Notations}

Besides the function $\displaystyle g(x,y)=\frac{-ic}{x-y}$ that
 enters in the definition of the $R$-matrix, we 
 also introduce $f(x,y)=1+g(x,y).$

\paragraph{Representations} are labelled by the triplet $(\lambda_1(u), \lambda_2(u), \lambda_3(u))$.
Associated to the representation \eqref{eq:Lvac} of the Lax operator, we have the functionals 
$$r_1(u)=\frac{\lambda_1(u)}{\lambda_2(u)}\equiv1\to1
\mb{and}
r_3(u)=\frac{\lambda_3(u)}{\lambda_2(u)}\equiv \left(\frac{1-\frac{iu\Delta}2}{1+\frac{iu\Delta}2}\right)^M\to e^{iuL}$$ 
where we have indicated (through $\to$) their value in the continuous limit.

%
\paragraph{We will use many sets of variables.} The notation will be as follows:
\begin{itemize}
\item{"bar"} always denotes {sets} of variables: $\bar w$, $\bla$, $\bmu$ etc..

\item{Individual elements} of the sets have {latin subscripts}: $w_j$, $u_k$,  etc..

\item{Subsets} of variables are denoted by {roman indices}: $\bla_{\so}$, $\bmu_{\rm iv}$, $\bar w_{\st}$, etc.

\item{Special case:} $\bar u_j=\bar u\setminus\{u_j\}$, $\bar w_k=\bar w\setminus\{w_k\}$, etc...
\end{itemize}

\paragraph{We use also shorthand notation for products of commuting operators / functions: }

\beano
&& r_3(\bla_{\st})=\prod_{\la_j\in\bla_{\st}} r_3(\la_j);\quad
T_{12}({\bla}) = \prod_{\la_j\in{\bla}} T_{12}(\la_j)
\\
&& g(v_k, \bar w)= \prod_{w_j\in\bar w} g(v_k, w_j) ;\quad
 f(\bla_{\st},\bla_{\so})=\prod_{\la_j\in\bla_{\st}}\prod_{\la_k\in\bla_{\so}} f(\la_j,\la_k),\quad etc..
\eeano

\section{Standard results from ABA}
\paragraph{Explicit expressions for Bethe vectors.}
There are different expressions for BVs, see \cite{BVs}. Here, we will focus on a particular one:
\beano
\mathbb{B}^{a,b}(\bla;\bmu) &=& \sum {\Izer_{a}({\bmu}_{\so}|\bla)}
\frac{f(\bmu_{\st},\bmu_{\so})}{f(\bmu,\bla)}\,
T_{13}({\bmu}_{\so})T_{23}({\bmu}_{\st})|0\rangle,
\eeano
where $\Izer_{a}({\bmu}|\bla)$ is the Izergin determinant.

The Bethe vectors are (right-)eigenvectors of the transfer matrix $t(w)$
$$
t(w)\,\mathbb{B}^{a,b}(\bla;\bmu) = \tau(w|\bla;\bmu) \,\mathbb{B}^{a,b}(\bla;\bmu) 
$$
when the {Bethe equations} are obeyed (then, the vectors are called {on-shell})
$$
\begin{cases}&r_1(u_j)=\frac{f(u_{j},\bla_{j})}{f(\bla_{j},u_j)}f(\bmu,u_j)\\
&j=1,2,...,a\end{cases}\ ; \qquad
\begin{cases}
&r_3(v_k)=\frac{f(\bmu_{k},v_k)}{f(v_k,\bmu_{k})}f(v_k,\bla)\\
&k=1,2,...,b.
\end{cases}
$$

In the same way, dual Bethe vectors $\mathbb{C}^{a,b}(\bla;\bmu)$ are left eigenvectors (when on-shell)
$$
\mathbb{C}^{a,b}(\bla;\bmu)\,t(w) = \tau(w|\bla;\bmu) \,\mathbb{C}^{a,b}(\bla;\bmu)\,.
$$

\paragraph{Global form factors.}
The form factors for $T_{ij}(u)$ were computed in \cite{twisted,FF}.
For {on-shell Bethe vectors} $\mathbb{C}^{a,b}(\blac;\bmuc)$ and $\mathbb{B}^{a,b}(\blab;\bmub)$
they have the following form:
\bea\label{eq:formfactor}
\mathcal{F}_{a,b}^{(i,j)}(z|\blac,\bmuc;\blab,\bmub) &=& 
\mathbb{C}^{a,b}(\blac;\bmuc) T_{ij}(z) \mathbb{B}^{a,b}(\blab;\bmub) \\
&=& 
\Big(\tau(z|\blac,\bmuc)-\tau(z|\blab,\bmub)\Big)\,\UF_{a,b}^{(i,j)}(\blac,\bmuc;\blab,\bmub).
\eea

$\UF_{a,b}^{(i,j)}(\blac,\bmuc;\blab,\bmub)$ is the \textsl{universal form factor.}
It  does not depend on the spectral parameter $z$, it is
 independent of the representation of $T(z)$ and admits a single determinant form 
  for the GL(3) invariant R-matrix.

However, $T(u)$ provides acces only to global operators: to get form factors of local operators, we need to consider a more refined model, that we present now.

\section{The composite model}
The composite model framework was introduced in \cite{IZ1984}.
In the lattice version, we consider $T(u)=T^{(2)}(u|m)\, T^{(1)}(u|m)$, where
\beq
T^{(2)}(u|m)=\,L(u|M)\cdots L(u|{m+1})\mb{and}
T^{(1)}(u|m)= \,L(u|m)\cdots L(u|1)\,.
\eeq
The integer $m\in [1,M[$ plays the role of the position $x$ in the continuous version, and
$T^{(j)}(u|m)$ are monodromy matrices for "shorter chains".

\paragraph{Vacuum expectation value.} The vacuum has a factorized form $|0\rangle=|0\rangle^{(1)}\,|0\rangle^{(2)}$
so that
\beano
T_{jj}^{(1)}(u|m)\,|0\rangle^{(1)}=\ell_j(u)\,|0\rangle^{(1)}\,,\quad j=1,2,3\,.
\eeano
For the Bose gas model, $\ell_1(u)=\ell_2(u)=1$ and 
$$\ell_3(u)=\Big(\frac{1+\frac{iu\Delta}2}{1-\frac{iu\Delta}2}\Big)^m\to e^{iux}.$$

\paragraph{Bethe vectors in composite model. \cite{nikita,compo1}} An explicit expression is given by
\beano
\mathbb{B}^{a,b}(\bla;\bmu)&=&
\sum \frac{\ell_{3}(\bmu_{\st})}{\ell_{1}(\bla_{\so})} f(\bla_{\so},\bla_{\st})f(\bmu_{\st},\bmu_{\so})f(\bmu_{\so},\bla_{\so})
\ \mathbb{B}_{a_{\so},b_{\so}}^{(1)}(\bla_{\so};\bmu_{\so})\ \mathbb{B}_{a_{\st},b_{\st}}^{(2)}(\bla_{\st};\bmu_{\st}),
\eeano
where $\mathbb{B}_{a_{\so},b_{\so}}^{(j)}(\bla;\bmu)$ are Bethe vectors for the partial monodromy matrices 
$T^{(j)}(u|m)$, $j=1,2$.

Note that the 'partial' BVs $\mathbb{B}_{a_{\so},b_{\so}}^{(j)}(\bla;\bmu)$, $j=1,2$, are off-shell
 even when the total BV $\mathbb{B}^{a,b}(\bla;\bmu)$ is on-shell.

\paragraph{Local form factors. \cite{compo2}} 
We are interested in computing
$\mathbb{C}^{a,b}(\blac;\bmuc)\, T_{ij}^{(1)}(u|m)\, \mathbb{B}^{a,b}(\blab;\bmub)$,
where the Bethe vectors are on-shell. It is a local form factor because of the position $m$.

More precisely, we want to compute 
{\beano
\sM_{a,b}^{(i,j)}(x|\blac,\bmuc;\blab,\bmub)=
\mathbb{C}^{a,b}(\blac;\bmuc)\, T^{(1)}_{ij}[0]\, \mathbb{B}^{a,b}(\blab;\bmub) 
\eeano}
where $T^{(1)}_{ij}[0]$ is the {zero mode (the Lie algebra part)} of $T_{ij}^{(1)}(u|m)$, that we detail below.

\section{The zero mode method}
\paragraph{The zero modes for the Bose gas} are defined as
$$
\begin{aligned}
&T_{ij}[0]=\lim_{|u|\to\infty}\frac uc(T_{ij}(u)-\delta_{ij}),
\quad i,j=1,2\quad
&T_{33}[0]=\lim_{|u|\to\infty}\frac uc(e^{-iLu}T_{33}(u)-1),\\
&T_{j3}[0]=\lim_{u\to -i\infty}\frac uc e^{-iLu}T_{j3}(u),\quad 
&T_{3j}[0]=\lim_{u\to i\infty}\frac uc T_{j3}(u),\quad j=1,2
\end{aligned}
$$
They generate an $SL(3)$ Lie algebra:
\beano
&&\big[ T_{ij}[0]\,,\, T_{kl}[0] \big] = \delta_{i,l}\,T_{kj}[0]-\delta_{j,k}\,T_{il}[0],\quad i,j,k,l=1,2,3\,,
\\
&&\big[ T_{ij}[0]\,,\, T_{kl}(z) \big] = \delta_{i,l}\,T_{kj}(z)-\delta_{j,k}\,T_{il}(z),\quad i,j,k,l=1,2,3\,.
\eeano
In particular $\big[ T_{ij}[0]\,,\, t(z) \big] =0$: they are a symmetry of the model.

\paragraph{Bethe vectors and zero modes.} Zero modes appear naturally in the expression of BVs:
\beano
&&T_{12}[0]\mathbb{B}^{a,b}(\bla;\bmu)=\lim_{|w|\to\infty} \tfrac wc\;\mathbb{B}^{a+1,b}(\{w,\bla\};\bmu), \qquad
\\
&&T_{23}[0]\mathbb{B}^{a,b}(\bla;\bmu)=\lim_{w\to-i\infty} e^{-iwL}\, \tfrac wc\;\mathbb{B}^{a,b+1}(\bla;\{w,\bmu\}),\qquad
\eeano
with similar expressions for dual Bethe vectors. 
Note that the limit $|w|\ \to\ \infty$ preserves the Bethe equations, i.e. the Bethe vector stays on-shell.

Moreover, for on-shell Bethe vectors:
\beano
&&T_{21}[0]\,\mathbb{B}^{a,b}(\bla;\bmu)=T_{32}[0]\,\mathbb{B}^{a,b}(\bla;\bmu)=T_{31}[0]\,\mathbb{B}^{a,b}(\bla;\bmu)=0
\\
&&\mathbb{C}^{a,b}(\bla;\bmu)\,T_{12}[0]=\mathbb{C}^{a,b}(\bla;\bmu)\,T_{23}[0]=\mathbb{C}^{a,b}(\bla;\bmu)\,T_{13}[0]=0,
\eeano
where the Bethe parameters $\bar u$ and $\bar v$ are supposed to be finite. 

\paragraph{The zero mode method} is a way to compute form factors using zero modes \cite{zeroM}. The main idea
is to use the Lie algebra symmetry generated by the zero modes and the highest weight property of (on-shell) Bethe vectors to obtain relations among the form factors \eqref{eq:formfactor}. As an example, let us show how to relate $\mathcal{F}_{a+1,b}^{(2,2)}(z|\blac,\bmuc;\{\blab,w\},\bmub)$ to $\mathcal{F}_{a,b}^{(1,2)}(z|\blac,\bmuc;\blab,\bmub)$:
\beano
&&\lim_{w\to\infty}\frac wc \mathcal{F}_{a+1,b}^{(2,2)}(z|\blac,\bmuc;\{\blab,w\},\bmub)
=
\mathbb{C}^{a,b}(\blac;\bmuc)\, T_{22}(z)\,  \lim_{w\to\infty}\frac wc \,\mathbb{B}^{a+1,b}(\{\blab,w\};\bmub) )\\
&&\qquad=
\mathbb{C}^{a,b}(\blac;\bmuc)\, T_{22}(z)\, T_{12}[0]\,\mathbb{B}^{a,b}(\blab;\bmub) )
= 
\mathbb{C}^{a,b}(\blac;\bmuc)\, \big[T_{22}(z)\,,\, T_{12}[0]\big]\,\mathbb{B}^{a,b}(\blab;\bmub) )\\
&&\qquad=
\mathbb{C}^{a,b}(\blac;\bmuc)\, T_{12}(z)\,\mathbb{B}^{a,b}(\blab;\bmub) )
=\mathcal{F}_{a,b}^{(1,2)}(z|\blac,\bmuc;\blab,\bmub).
\eeano
More examples of such relations can be found in \cite{zeroM}.

\paragraph{Zero mode in the composite model.} We can defined 'local' zero modes, through the composite model:
$$
\begin{aligned}
&T^{(1)}_{ij}[0]=\lim_{|u|\to\infty}\frac uc(T^{(1)}_{ij}(u|x)-\delta_{ij})
=-\int_0^x \psi_i^\dagger(y)\psi_j(y)\,dy, 
\quad i,j=1,2
\\
&T^{(1)}_{j3}[0]=\lim_{u\to -i\infty}\frac uce^{-ixu}T^{(1)}_{j3}(u|x)=-\frac1{\sqrt{c}} \psi_j(x),
\quad j=1,2
\\
&T^{(1)}_{3j}[0]=\lim_{u\to i\infty}\frac ucT^{(1)}_{3j}(u|x)=-\frac1{\sqrt{c}} \psi_j^\dagger(x),
\quad j=1,2
\\
&\big[ T^{(1)}_{ij}[0]\,,\, T_{kl}[0] \big] = \delta_{i,l}\,T^{(1)}_{kj}[0]-\delta_{j,k}\,T^{(1)}_{il}[0],\quad i,j,k,l=1,,2,3.
\end{aligned}
$$
The zero mode method applied to local zero modes
allows to relate the different local form factors $\sM_{a,b}^{(i,j)}(x|\blac,\bmuc;\blab,\bmub)$. Then, 
we need to compute only one local form factor. It is done using the twisted transfer matrix method.

\section{The twisted transfer matrix method}
Still associated to the monodromy matrix $T(u)$, one can introduce a twisted version of the transfer matrix
$$
t_{\bar\beta}(z) = \sum_{j=1}^3 e^{\beta_j}\,T_{jj}(z)=tr_0\big(M_{\bar\beta}\,T(u)\big)\,,\quad 
M_{\bar\beta}=\mbox{diag}\left(e^{\beta_1}\,,\,e^{\beta_2}\,,\,e^{\beta_3}\right)\,,\quad
\beta_j\in\CC
$$
that is integrable. Twisted BVs (constructed from ABA) can be associated to this twisted transfer matrix. 
They are eigenvectors of $t_{\bar\beta}(z)$
$$
t_{\bar\beta}(w)\,\mathbb{B}^{a,b}_{\bar\beta}(\bla;\bmu) = \tau_{\bar\beta}(w|\bla;\bmu) \,\mathbb{B}_{\bar\beta}^{a,b}(\bla;\bmu) 
$$
when the {Bethe equations} are obeyed 
$$
{r_1(u_j)=e^{\beta_1-\beta_2}\,\frac{f(u_{j},\bla_{j})}{f(\bla_{j},u_j)}f(\bmu,u_j),\quad
r_3(v_k)=e^{\beta_2-\beta_3}\,\frac{f(\bmu_{k},v_k)}{f(v_k,\bmu_{k})}f(v_k,\bla).}
$$
Accordingly, the twisted scalar product
$
\mathcal{S}_{(\bar\beta)}^{a,b}=\mathbb{C}^{a,b}_{\bar\beta}(\blac;\bmuc)\,\mathbb{B}^{a,b}(\blab;\bmub)
$
can be computed. It is known as a single determinant \cite{twisted}.

\paragraph{The twisted transfer matrix method} allows to compute diagonal form factors for the untwisted transfer matrix, starting from scalar products associated to the twisted one.
The generating functional for such form factors is given by
\beano
G_{(\bar\beta)}^{a,b} &=&
\mathbb{C}_{(\bar\beta)}^{a,b}(\blac;\bmuc)\,e^{Q_{\bar\beta}}\,\mathbb{B}^{a,b}(\blab;\bmub)
= e^{\beta_1\ell_1[0]+\beta_3\ell_3[0]}\
\frac{\ell_1(\blac)\ell_3(\bmub)}{\ell_1(\blab)\ell_3(\bmuc)}\;\mathcal{S}_{(\bar\beta)}^{a,b}\\
Q_{\bar\beta} &=& \sum_{j=1}^3 \beta_j\,T_{jj}^{(1)}[0]\,,\quad 
\beta_j\in\CC\,.
\eeano

The generating functional allows to compute the diagonal form factor
$$
\mathbb{C}^{a,b}(\blac;\bmuc)T^{(1)}_{ii}[0]\mathbb{B}^{a,b}(\blab;\bmub)=
\left(\frac{\ell_1(\blac)\ell_3(\bmub)}{\ell_1(\blab)\ell_3(\bmuc)}-1\right)\,
\frac{d}{d\beta_i}\mathcal{S}_{(\bar\beta)}^{a,b}\Bigr|_{\bar\beta=0}.
$$

\section{Results}
Gathering the different results exposed above, one can compute the desired form factors \cite{bose}:
\beano
&&\mathbb{C}_{a,b}(\bla;\bmu)\;{\psi_j^\dagger(x)\psi_j(x)}\;\mathbb{B}_{a,b}(\bla;\bmu)=
i\sum_{k=1}^b\frac{dv_k(\bar\beta)}{d\beta_j}
\Bigr|_{\bar\beta=0}\,\|\mathbb{B}_{a,b}(\bla;\bmu)\|^2, \qquad j=1,2,
\eeano

\beano
&& \mathbb{C}_{a',b}(\blac;\bmuc)\,{\psi_i^\dagger(x)\psi_j(x)}\,\mathbb{B}_{a,b}(\blab;\bmub)
=-i\mathcal{P}(\bmub,\bmuc)
\;e^{ix\mathcal{P}(\bmub,\bmuc)}\ \UF_{a,b}^{(i,j)}(\blac,\bmuc;\blab,\bmub),
\\
&&\qquad a'=a+j-i \quad \{\blac;\bmuc\}\neq\{\blab;\bmub\}
\\
&&\mathbb{C}_{a-2+k,b-1}(\blac;\bmuc)\,{\psi_k(x)}\,\mathbb{B}_{a,b}(\blab;\bmub)
= i\sqrt{c}\;e^{ix\mathcal{P}(\bmub,\bmuc)} \UF_{a,b}^{(3,k)}(\blac,\bmuc;\blab,\bmub),
\\
&&\mathbb{C}_{a+2-k,b+1}(\blac;\bmuc)\,{\psi^\dagger_k(x)}\,\mathbb{B}_{a,b}(\blab;\bmub)
= i\sqrt{c}\;e^{ix\mathcal{P}(\bmub,\bmuc)} {\UF}_{a,b}^{(k,3)}(\blac,\bmuc;\blab,\bmub).
\eeano
Above, ${\UF}_{a,b}^{(k,l)}(\blac,\bmuc;\blab,\bmub)$ are the universal form factors (that are known, see \cite{zeroM}), and
 $$\mathcal{P}(\bmub,\bmuc)=\sum_{i=1}^b\mub_i-\sum_{i=1}^{b'}\muc_i\,.$$

\section{Perspectives}

The present results will be used to calculate the mean values 
$\langle \psi^\dagger_j(x)\psi_j(x) \rangle$, $\langle \psi^\dagger_j(x)\rangle$ and $\langle \psi_j(x) \rangle$, 
for applications to condensed matter experiments.

We also obtained a
conjecture for the form factors of local operator in $gl(N)$ composite models:
\beano
\mathbb{C}_{\bar b}(\bar s)\;T^{(1)}_{ij}[0]\;\mathbb{B}_{\bar a}(\bar t) &=&
\left(\prod_{k=1}^{N-1}\frac{\alpha_k(\bar s^{k})}{\alpha_k(\bar t^{k})}-1\right)\UF_{\bar a}^{(i,j)}(\bar s;\bar t),
\quad \mbox{ for}\quad \bar s \neq\bar t
\\
\label{FF-Tee-2-GLN}
\mathbb{C}_{\bar a}(\bar t)\;T^{(1)}_{jj}[0]\;\mathbb{B}_{\bar a}(\bar t) &=&
\left( \lambda^{(1)}_{j}[0]+\sum_{k=1}^{N-1}\frac{d}{d\kappa_j}
\log\alpha_k(\bar t^{k}(\bar\kappa))
\Bigr|_{\bar\kappa=1}\right)\|\mathbb{B}_{\bar a}(\bar t)\|^2.\quad
\eeano
with $\displaystyle \alpha_j(u)=\frac{\lambda_{j}^{(1)}(u)}{\lambda_{j+1}^{(1)}(u)}$ ;
$T^{(1)}_{jj}(u)\,|0\rangle^{(1)}= \lambda_{j}^{(1)}(u)\,|0\rangle^{(1)},$\quad
 $j=1,\dots,N-1.$
 Note however that we still need a determinant form for the scalar product and/or $\UF_{\bar a}^{(i,j)}(\bar s;\bar t)$.

Finally, the superalgebra case is investigated: work is in progress for the rational $R$-matrix, for the computation 
of the Bethe vectors, scalar products and form factors of the model.


\end{document}